\documentstyle[psfig,gcdv]{article}
\def\reference{\parskip 0pt\par\noindent\hangindent 0.5 truecm}

\def\spose#1{\hbox to 0pt{#1\hss}}
\def\simlt{\mathrel{\spose{\lower 3pt\hbox{$\mathchar"218$}}
     \raise 2.0pt\hbox{$\mathchar"13C$}}}
\def\simgt{\mathrel{\spose{\lower 3pt\hbox{$\mathchar"218$}}
     \raise 2.0pt\hbox{$\mathchar"13E$}}}

\begin{document}

\small
\shorttitle{Tidal Debris from Omega Cen's Progenitor Galaxy}
\shortauthor{M.\ Chiba \& A.\ Mizutani}
%
%
\title{\large \bf
On the Kinematics of Tidal Debris from Omega Cen's
Progenitor Galaxy}

\author{\small 
 Masashi Chiba$^{1}$ and
 Arihiro Mizutani$^{2}$
} 

\date{}
\twocolumn[
\maketitle
\vspace{-20pt}
\small
{\center
$^1$ Astronomical Institute, Tohoku University, Sendai 980-8578, Japan\\
$^2$ National Astronomical Observatory, Mitaka, Tokyo 181-8588, Japan\\[3mm]
}

%
\begin{center}
{\bfseries Abstract}
\end{center}
\begin{quotation}
\begin{small}
\vspace{-5pt}
The kinematic properties of tidal debris from an orbiting Galactic satellite
is presented, on the assumption that its central part once contained the
most massive Galactic globular cluster, $\omega$ Cen. We simulate dynamical
evolution of a satellite galaxy that follows the present-day
and likely past orbits of $\omega$ Cen and analyze the kinematic nature
of their tidal debris and randomly generated Galactic stars
comprising spheroidal halo and flat disk components. It is found that the debris
stars show a retrograde rotation at $\sim -100$ km~s$^{-1}$, which may
accord with a recently discovered stellar stream at radial velocity of
$\sim 300$ km~s$^{-1}$ towards the 
Galactic longitude of $\sim 270^\circ$. These stars also contribute, only in
part, to a reported retrograde motion of the outer halo in the North Galactic
Pole direction, without significantly modifying local halo kinematics near
the Sun.
The prospects for future debris searches and the implications for the early
evolution of the Galaxy are briefly presented.
\end{small}
\end{quotation}
]


\bigskip

\section{Introduction}

It is well known that the most massive globular cluster in the Milky Way,
$\omega$ Cen, shows unique properties in its metallicity content, internal
kinematics, and structure. For instance, $\omega$ Cen shows a wide spread
in metallicity unlike other Galactic globular clusters (e.g. Norris, Freeman,
\& Mighell 1996): its metallicity distribution is peaked at
[Fe/H]$\simeq -1.6$, along with a second smaller peak at [Fe/H]$\simeq -1.2$
and a long tail extending up to [Fe/H]$\simeq -0.5$. Also, the metal-rich stars
in $\omega$ Cen are largely enhanced in $s$-process elements relative to those
in globular clusters and field stars with similar metallicities (e.g. Norris
\& Da~Costa 1995). This suggests that the ejecta from low-mass,
asymptotic giant branch (AGB) stars had to be retained and incorporated
into the next-generation stars. However, although $\omega$ Cen is most massive
($5\times 10^6$ M$_\odot$),
it is unable to retain the AGB ejecta, as shown by Gnedin et al. (2002).
Thus, an isolated formation of $\omega$ Cen is unlikely, because the
enriched gas would easily be lost by encountering the Galactic disk. The most
viable explanation for the uniqueness of $\omega$ Cen is that it was once
the dense nucleus of a dwarf galaxy, i.e., a nucleated dwarf (Freeman 1993).
A gravitational potential of progenitor's dark matter
would help retaining the enriched gas and let the cluster being self-enriched
at least over a few Gigayears.

In this contribution, we pursue the possible kinematical evidence for
the existence of such a dwarf galaxy, which was already disrupted in the past.
Dinescu (2002) first investigated this issue, by examining the possible signature
of the progenitor's tidal debris among nearby metal-poor stars in the catalog
of Beers et al. (2000, B00).
She identified a group of stars with $-2.0<$[Fe/H]$\le-1.5$, which
departs from the characteristics of the inner Galactic halo but has
retrograde orbits similar to $\omega$ Cen. Her simplified disruption model of
the progenitor galaxy demonstrated that trailing tidal debris, having
orbital characteristics similar to the cluster, can be found in the solar
neighborhood, although the concrete spatial distribution and kinematics
of the debris stars remain yet unclear.

Dinescu's work motivates us to undertake a more refined approach to the issue,
i.e., to conduct an N-body simulation for the tidal
disruption of $\omega$ Cen's progenitor galaxy (Mizutani et al. 2003).
We obtain the characteristic structure and kinematics of its debris stars
and compare with various observations showing signatures of recent merging
events in the Milky Way (Gilmore, Wyse, \& Norris 2002, GWN;
Kinman et al. 2003, K03; Chiba \& Beers 2000, CB).
In particular, we show that a recently identified stream of stars at
radial velocity of $\sim 300$ km~s$^{-1}$ (GWN) is a natural
outcome of the current disruption model, without significantly modifying
local halo kinematics near the Sun.

\section{Model of a Progenitor Galaxy}

For the purpose of enlightening the characteristics of tidal debris, we first
assume a fixed external gravitational potential (representing the Milky Way)
for the dynamical evolution of an orbiting dwarf galaxy.
The potential consists of a spherical Hernquist bulge
$\Phi_b(r)$, a Miyamoto-Nagai disk $\Phi_d (R,z)$, and a logarithmic
dark halo $\Phi_h(r)$, where $r$ is the Galactocentric distance and
$(R,z)$ denote cylindrical coordinates. Each is given as,
$\Phi_b (r) = - GM_b/(r+a)$, $\Phi_d(R,z) =
- GM_d/\sqrt{R^2+(b+\sqrt{z^2+c^2})^2}$,
and $\Phi_h(r) = v_h^2 /2 \ln(r^2+d^2)$,
where $M_b=3.4\times 10^{10}$ M$_\odot$, $a=0.7$ kpc,
$M_d=10^{11}$ M$_\odot$, $b=6.5$ kpc, $c=0.26$ kpc, $v_h=186$ km~s$^{-1}$,
and $d=12$ kpc. This choice yields a circular velocity
of 228 km~s$^{-1}$ at the solar circle of $R_\odot=8$ kpc and a flat
rotation curve outside $R_\odot$.
On our adoption of a logarithmic dark halo, we note that while a dark halo
may be modeled by the universal profile proposed by Navarro, Frenk, and White
(NFW profile), such a profile as obtained by N-body simulations can be modified
by the inclusion of baryonic gas, mostly in the inner part of a galaxy where
its density is increased owing to baryonic infall, in such a manner that the
density profile may be changed to an isothermal one (e.g. Barnes 1987).
It is also noted that the adoption of the NFW profile does not affect our results,
as the inner density profile in concern is dominated by the bulge and disk
components. A dwarf galaxy orbiting in this galactic potential is
then represented by self-gravitating particles following a King
model, where the central density, central velocity dispersion, and core radius
are given as 0.3 $M_\odot$~pc$^{-3}$, 18.1 km~s$^{-1}$, and 0.56 kpc,
respectively. In addition, a particle with the mass of $5 \times 10^6$ M$_\odot$
representing $\omega$ Cen is placed at the center of the
galaxy. This setting yields the total mass of the system as
$M_{tot}=5.79 \times 10^8$ M$_\odot$. On the other hand, the total mass of stars
in this hypothetical galaxy is estimated from the mean metallicity of
stars in $\omega$ Cen ($\langle$[Fe/H]$\rangle\sim-1.6$), combined
with the metallicity-luminosity relation for the Local Group dwarfs
(C\^{o}t\'{e} et al. 2000) and the mass-to-light ratio (assuming $M/L \sim 4$),
yielding $M_{stars}\sim 10^7$ M$_\odot$. Thus, the mass of the simulated galaxy
is largely dominated by dark matter. The model galaxy is
represented by a collection of $10^4$ particles and the self-gravity is
calculated in terms of a multiple expansion of the internal potential to
fourth order (White 1983; Zaritsky \& White 1988).

This model dwarf is disrupted by Galactic tides in the course of its orbital
motion, whereas its dense core is expected to survive and follow
$\omega$ Cen's orbit.
We examine two representative orbits for the progenitor, models 1
and 2: model 1 follows the current orbit of $\omega$ Cen, whereas for model 2,
we calculate an orbit back to the past over $\sim 2$ Gyr from its current
position and velocity taking into account dynamical friction
and then set a progenitor
galaxy on its non-decaying orbit. These two models provide us with satisfactory
information on the generic properties of a tidally disrupted progenitor and
we postulate that the realistic nature of their debris is midway between
these model predictions.
We calculate $\omega$ Cen's orbit, based on the current distance
$D=5.3 \pm 0.5$ kpc
from the Sun, position $(l,b)=(309^\circ,15^\circ)$, proper motion
$(\mu_\alpha \cos\delta,\mu_\delta) =(-5.08\pm0.35, -3.57\pm0.34)$
mas~yr$^{-1}$, and heliocentric radial velocity $v_{los} = 232.5 \pm 0.7$
km~s$^{-1}$ (Dinescu, Girard, \& van Altena 1999, DGvA).
This orbit for model 1 is characterized by frequent disk crossings with a period
of $\tau_{orb} = 0.8 \times 10^8$ yr, retrograde motion, and apo and pericentric
distances $(r_{apo},r_{peri}) = (6.4, 1.1)$ kpc. For model 2,
we obtain $\tau_{orb} = 1.5 \times 10^8$ yr and
$(r_{apo},r_{peri}) = (11.3, 3.0)$ kpc. 
In both experiments, we place
a progenitor galaxy at apocenter to maximize its survival chances.

We plot the spatial distribution of the tidally disrupted debris in Figure 1.
Upper (middle) panel shows model 1 (model 2) after the 1.37 (1.86) Gyr orbital
excursion of the progenitor galaxy. Lower panel shows the orbit of the galaxy
center.
A rosette-like feature of the debris becomes steady after about eight orbital
periods. Model 1 results in more compact distribution
than model 2, which reflects the difference in orbital radii.
Figure 2 shows the velocity distributions of the debris particles in
cylindrical coordinates $(v_R,v_\phi,v_z)$.
As is evident, both models provide essentially the same debris kinematics:
most notorious is a sharply peaked $v_\phi$ distribution at
$\sim -100$ km~s$^{-1}$, arising from a retrograde orbit of a progenitor.
These kinematics
suggest that the difference in model 1 and 2 resides only in the spatial
extent of the debris.

\begin{figure}[htb]
 \begin{center}
\begin{tabular}{c}
   \psfig{file=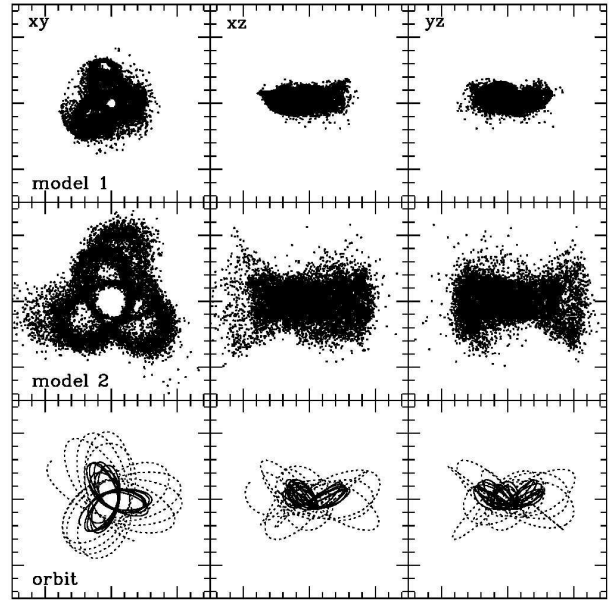,width=8.0cm} \\
 \end{tabular}
\caption{
Upper (middle) panel shows the spatial distribution of the tidally disrupted
debris for model 1 (model 2) after the 1.37 (1.86) Gyr orbital excursion of
$\omega$ Cen's progenitor galaxy. Lower panel shows the orbit of the galaxy
center for model 1 (solid line) and 2 (dotted line). The plots are projected
onto three orthogonal planes, where the Sun is located at $x=-8$ kpc and $xy$
corresponds to the disk plane. A frame measures 15 kpc on a side of each panel.
The current position of $\omega$ Cen is at $(x,y,z)= (-4.8,-4.0,1.4)$ kpc.}
 \end{center}
\end{figure}

\begin{figure}[htb]
 \begin{center}
 \begin{tabular}{c}
   \psfig{file=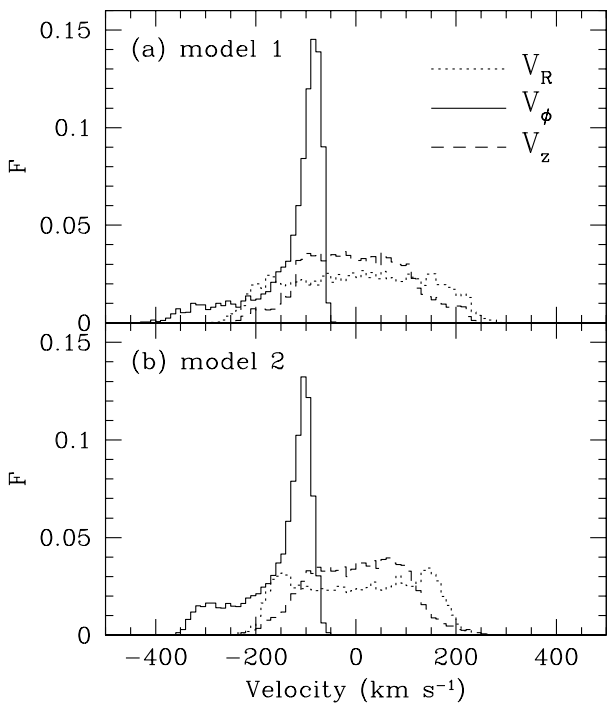,width=8.0cm} 
 \end{tabular}
\caption{
Velocity distribution of the debris particles in cylindrical coordinates,
$v_R$ (dotted), $v_\phi$ (solid), $v_z$ (dashed), for
model 1 (a) and 2 (b).
}
 \end{center}
\end{figure}

\section{Kinematics of Stream and Galactic Stars}

\subsection{Method}

We analyze the kinematics of
both the simulated debris and other Galactic stars
generated randomly by a Monte Carlo method. The metal-poor halo
is modeled as a flattened spheroid $\rho \propto
(R^2 + z^2/q^2)^{-3.5/2}$, where $q$ is an axis ratio ranging 0.55-0.7,
anisotropic velocity ellipsoid $(\sigma_R,\sigma_\phi,\sigma_z)=(154,121,96)$
km~s$^{-1}$, and small mean rotation $\langle v_\phi \rangle= 24$ km~s$^{-1}$,
as found for halo stars with [Fe/H]$<-2$ near the Sun (CB).
Thin and thick disks are modeled as $\rho \propto
\exp(-R/R_d) \sec^2 (z/z_d)$, where $R_d=3.5$ kpc and $z_d=0.3$ (1)
kpc for thin (thick) disk. Both disks
rotate at 200 km~s$^{-1}$, having velocity ellipsoids of
$(34,25,20)$ km~s$^{-1}$ and $(46,50,35)$ km~s$^{-1}$ for thin and thick
disks, respectively (CB). The relative fraction of each
component is fixed using observed local densities near the Sun,
in such a manner that the halo and thick-disk densities at $D<1$ kpc are 0.2~\%
and 2~\% of the thin-disk density, respectively (Yamagata \& Yoshii 1992).

In our model of $\omega$ Cen's progenitor galaxy,
the self-gravitating particles represent both stars and dark matter particles.
We note that the correct conversion of the mass of the simulated system into
that of the presumed stellar system alone remains yet uncertain, because
the correct $M/L$ ratio of stars as well as the amount of dark matter
in the progenitor galaxy is unavailable. As a useful model parameter to
incorporate this ambiguity for the current kinematic analysis, we set a
quantity $f$ as the fraction of the debris particles relative to halo stars
near the Sun, so that the normalization of the halo density is fixed for
the given number of the neighbor debris and $f$.

A typical value of $f$ for the conversion of the simulated particles to
the stars is estimated in the following manner. Model 1 (model 2) yields
21 (74) particles at $D < 2$ kpc, giving the mass density of
$\rho_g = 0.4 (1.3) \times 10^{-4}$ M$_\odot$~pc$^{-3}$ near
the Sun, whereas the total mass density and
metal-poor halo density have been derived as
$8 \times 10^{-3}$ M$_\odot$~pc$^{-3}$ (Gates, Gyuk, \& Turner 1995) and
$6.4 \times 10^{-5}$ M$_\odot$~pc$^{-3}$ (Gould, Flynn, \& Bahcall 1998),
respectively.
Then, if the debris stars (with $M_{stars} \sim 10^7$ M$_\odot$)
are distributed in the same manner as the simulated particles (with
$M_{tot} =5.74 \times 10^8$ M$_\odot$), which would be a reasonable
approximation in view of the dissipationless nature of stars,
the mass density of the debris stars in the solar neighborhood can be
estimated as $(M_{stars}/M_{tot}) \rho_g = O(10^{-6})$
M$_\odot$~pc$^{-3}$, which is about 1~\% of the halo density.
Thus, $f$, defined here at $D<2$ kpc, is expected to be of order of
a few percents.

\begin{figure}[htb]
 \begin{center}
 \begin{tabular}{c}
   \psfig{file=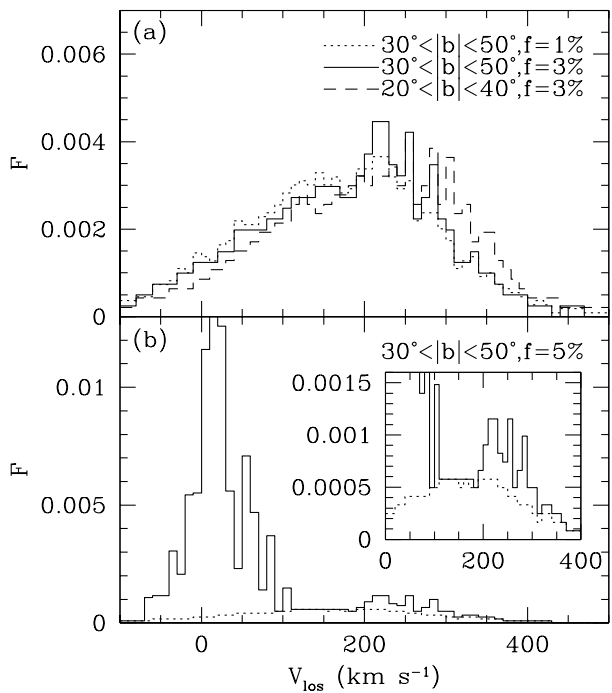,width=8.0cm} \\
 \end{tabular}
\caption{
(a) Distribution of the heliocentric radial velocities in the direction of
the GWN survey, for the debris stars of model 2
and randomly generated halo stars with $q=0.7$.
We select the stars at $1<D<5$ kpc and $260^\circ<l<280^\circ$
in the fields of $30^\circ<|b|<50^\circ$ (with $f=1$~\% and 3~\% for
dotted and solid histograms, respectively) and $20^\circ<|b|<40^\circ$
(with $f=3$~\%: dashed histogram).
(b) The same as (a) but incorporating the randomly generated disk
stars as well for $30^\circ<|b|<50^\circ$ with $f=5$~\% (solid histogram).
Dotted histogram denotes the contribution from the metal-poor halo alone.
}
 \end{center}
\end{figure}

\subsection{Radial velocity distribution at $l \simeq 270^\circ$}

GWN recently reported their spectroscopic survey of
$\sim 2000$ F/G stars down to $V = 19.5$ mag, in the direction
against Galactic rotation $(l,b) = (270^\circ , -45^\circ)$ and
$(270^\circ , +33^\circ)$,
where radial velocities in combination of distances largely
reflect orbital angular momentum. Their calibration of the stars at distances
of a few kpc from the Sun leads to the discovery of two stellar streams
at $v_{los}\sim 100$ km~s$^{-1}$ and $\sim 300$ km~s$^{-1}$, which are not
explained by any existing Galactic components (see their Fig.2 and 3).
While the stream at $v_{los} \sim 100$ km~$^{-1}$ was reproduced by
their model of a merging satellite in prograde rotation,
the stream at $v_{los}\sim 300$ km~$^{-1}$ remains yet unexplained.

Figure 3a shows the $v_{los}$ distribution for the
simulated debris of model 2 and halo stars of $q=0.7$ (i.e. without disks)
at $1<D<5$ kpc, $260^\circ<l<280^\circ$, and two fields for $b$.
Figure 3b shows when disk stars are incorporated. As is evident,
the debris stars from $\omega$ Cen's progenitor galaxy form a local peak
at $v_{los}\sim 300$ km~s$^{-1}$, which is provided by many stars having
$v_\phi \sim -100$ km~$s^{-1}$ (Fig. 2). This is in good
agreement with the $v_{los}\sim 300$ km~s$^{-1}$ stream discovered by GWN.
A more flattened halo than the case $q=0.7$, which is possible within a
range of $0.55 \le q \le 0.7$ (CB), yields a higher peak,
since the density contrast of
the debris relative to halo is made higher in this survey region.
We also make a plot for model 1 but do not present it here as it is
essentially the same as Figure 3 for model 2.
This reflects that model 1 shows the same velocity distribution as model 2
(Fig. 2), although to attain the same peak height at $\sim 300$ km~s$^{-1}$,
$f$ be a few factor larger and the selected range of $l$ be a few degree
higher than the respective values in model 2, because of less number of
debris particles near the Sun. This rule applies to other considerations
below as well.

\subsection{Kinematics at the North Galactic Pole}

The halo kinematics at the North Galactic Pole (NGP) deserve
special attention.
Majewski (1992) suggested that the outer halo at the NGP shows
a retrograde rotation $\langle v_\phi \rangle \sim -55$ km~s$^{-1}$ at $z>4$ kpc.
Also, K03 reported from their analysis of RR Lyrae
and blue horizontal branch stars that the halo at $2<z<12$ kpc
shows a retrograde rotation at $\langle v_\phi \rangle \sim -65$ km~s$^{-1}$.
On the other hand, halo stars near the Sun 
show no significant retrograde rotation (CB).

For the purpose of elucidating the effect of the debris stars on
the NGP kinematics, we select
the debris of model 2 and randomly generated stars at $b>70^\circ$ and
$2<D<5$ kpc (resembling K03's selection). Since the observational determination
of full space velocities involves rather
inaccurate information of proper motions compared to radial velocities,
we convolve the velocity distribution of stars with a Gaussian
distribution for velocity errors, having 1~$\sigma$ of a typical
30 km~s$^{-1}$ error. Figure 4 shows the velocity distributions at various
values of $f$. It follows that the debris stars provide a non-Gaussian feature
in the velocity distribution. 
The $v_\phi$ distribution shows an extra peak at $\sim -100$ km~$s^{-1}$,
whereby the debris stars decrease $\langle v_\phi \rangle$ from that of a pure
halo sample. However,
the change of $\langle v_\phi \rangle$ by the inclusion of the debris stars
with $f=5$~\% amounts to only $-19$ ($-14$) km~s$^{-1}$ for $q=0.55$ ($0.7$),
which are still insufficient for explaining
the reported $\langle v_\phi \rangle = -35 \sim -75$ km~s$^{-1}$.
For the $v_R$ and $v_z$ distributions, their means at the cases of $q=0.7$ and
$f=5$~\% are slightly increased to $\langle v_R \rangle = 7$ km~s$^{-1}$ and
$\langle v_z \rangle = 3$ km~s$^{-1}$, respectively, whereas
the velocity dispersions are decreased by the amounts of 6 km~s$^{-1}$ and
5 km~s$^{-1}$, respectively, when the debris stars are considered.
Note that if we extend the selection of the stars at higher $z$ or instead
consider model 1, the changes of $\langle v_\phi \rangle$ become smaller than
the above mentioned values, because there are no debris stars in our current
model (Fig. 1). Thus, it is safe to conclude that the debris stars
contribute only in part to a reported retrograde motion at the NGP.

\begin{figure}[htb]
 \begin{center}
 \begin{tabular}{c}
   \psfig{file=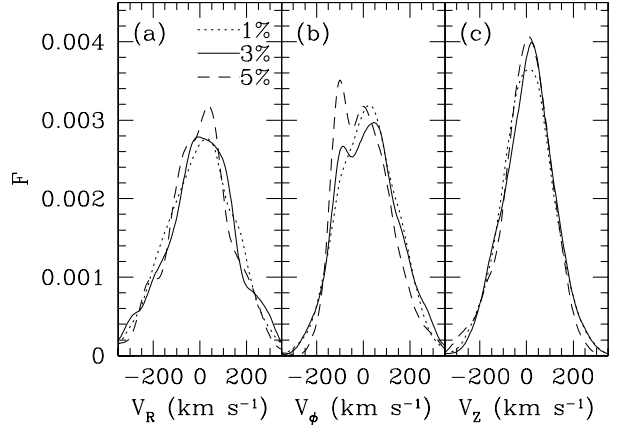,width=8.0cm} \\
 \end{tabular}
\caption{
Velocity distribution of the simulated stars of model 2 in the North Galactic
Pole direction. In this plot, we select the stars at $b>70^\circ$ and $2<D<5$
kpc and convolve their derived velocity distribution with a Gaussian error
distribution having 1~$\sigma$ of 30 km~s$^{-1}$. The axis ratio of the
metal-poor halo is set as $q = 0.7$. Dotted, solid, and dashed lines denote
$f=1$~\%, 3~\%, and 5~\%, respectively.
}
 \end{center}
\end{figure}

\subsection{Local halo kinematics}

We select the simulated stars at the cylindrical coordinates of $6.5<R<9.5$ kpc
and $z<4$~kpc and at the distance from the Sun of $D<4$ kpc (as was drawn by CB),
convolve the velocities
with a Gaussian error distribution of $1~\sigma = 30$ km~s$^{-1}$, and
compare with the corresponding stars with [Fe/H]$\le-2$ in B00.
It follows that the non-Gaussian feature of the velocity
distribution is much weaker than that at the
NGP: the change of $\langle v_\phi \rangle$ for $f=5$~\% is only
$-9$ ($-10$) km~s$^{-1}$ in model 1 (model 2). This is due to the
characteristic debris distribution, where the number
density near $z=0$ is smaller than at high $|z|$ (Fig. 1).

\section{Discussion}

Our simple model of an orbiting dwarf galaxy that once contained $\omega$ Cen
predicts a sequence of tidal streams in retrograde rotation and their existence
can be imprinted in kinematics of nearby stars, especially in the direction
against Galactic rotation (GWN) and at the NGP (K03), while local halo
kinematics remain unchanged. The simulated streams are mostly distributed
inside the solar circle, as suggested from the current orbital motion of
$\omega$ Cen (DGvA; Dinescu 2002).
In contrast to Sgr dwarf galaxy having polar orbit
(Ibata et al. 1997), the orbit of $\omega$ Cen's progenitor galaxy is
largely affected by a non-spherical disk potential, where the orbital plane
exhibits precession with respect to the Galactic Pole, causing self-crossing
of tidal streams in the disk region (Fig. 1). The projection of the orbit
perpendicular to the disk plane shows an 'X'-like feature, thereby leaving
denser streams at high $|z|$ than at low $|z|$ for a given radius.
These characteristic spatial distributions of the debris stars give rise
to more significant effects of the debris at the NGP than
in the solar neighborhood, as shown here, although the current simulation
failed to reproduce the reported largely retrograde rotation at NGP from
the $\omega$ Cen debris alone; perhaps, other, yet unknown halo substructures
must be considered to reproduce the observations.

Existing kinematic studies of Galactic stars to search for a signature
of $\omega$ Cen's progenitor galaxy are yet confined to nearby stars,
where the significance of the debris streams is modest, as shown here.
Searches of stars inside the solar circle are more encouraging (Fig. 1),
in particular in the directions of $l \sim 320^\circ$ and $l \sim 50^\circ$,
where we expect the presence of high-velocity streams at
$v_{los} = 200 \sim 300$ km~s$^{-1}$ and $-400 \sim -300$ km~s$^{-1}$,
respectively. Future radial velocity surveys of these fields including the
sample of the Sloan Digital Sky Survey or planned Radial Velocity
Experiment are worth exploring in this respect. Also, detailed
abundance studies of candidate stream stars will be intriguing, because such
stars may exhibit different abundance patterns from field halo stars,
as found in dwarf galaxies (Shetrone, C\^{o}t\'{e}, \& Sargent 2001).

In this work, we adopt a fixed external gravitational potential
for the calculation of an orbiting dwarf galaxy and its dynamical evolution,
thereby neglecting dynamical friction against the satellite. However,
to set more refined limits on its dynamical history, it is required to
fully take into account frictional effect on the orbit as well as dynamical
feedback of a satellite on the structure of the Galactic disk.
For instance, Tsuchiya, Dinescu, and Korchagin (2003) reported their
numerical models for a satellite with strong orbital decay
and succeeded to reproduce the current orbit of $\omega$ Cen from the
launch of the progenitor satellite at 58 kpc from the Galactic center.
They also showed that the debris particles at 3 Gyr are already smeared out
and distributed in a flattened disk without having significant stream-like
features as obtained here. It is noted that whether or not stream-like
features survive by the current epoch depends on when and how a satellite
galaxy is disrupted by Galactic tides, or in other words, the observational
information on such features and the comparison with simulation results
are useful for placing important constraints on when a satellite merging occurs.
For this purpose, we are currently undergoing full N-body simulations of several
host-satellite systems using GRAPE5, where both host and satellite galaxies
are represented by lively dark halos and stellar components.
Our goals with the use of GRAPE5 are to set tight
limits on the initial total mass, internal mass distribution, and orbital motion
of a progenitor galaxy, as well as the timing of merging with the Galactic
disk, based on the comparison between the observation (e.g., RAVE)
and simulation results for stream-like structures in the Milky Way.
More details will be reported elsewhere (Mizutani \& Chiba, in preparation).

\section*{Acknowledgements}

M.C. thanks Kenji Bekki and Tim Beers for useful discussions.
We are grateful to anonymous referees for invaluable comments on
the manuscript.

\section*{References}
\reference Barnes, J. E. 1987, 
in Nearly normal galaxies: From the Planck time to the present, ed. S. Faber,
(New York: Springer-Verlag), 154
\reference Beers, T. C., Chiba, M., Yoshii, Y., Platais, I.,
Hanson, R. B., Fuchs, B., \& Rossi, S. 2000, AJ, 119, 2866 (B00)
\reference Chiba, M., \& Beers, T. C. 2000, AJ, 119, 2843 (CB)
\reference C\^{o}t\'{e}, P., Marzke, R. O., West, M. J., \& Minniti, D.
2000, ApJ, 533, 869
\reference Dinescu, D. I. 2002, in ASP Conf. Ser. 265, Omega Centauri: A
Unique Window into Astrophysics, ed. F. van Leeuwen, J. D. Hughes, \&
G. Piotto (San Francisco: ASP), 365
\reference Dinescu, D. I., Girard T. M., \& van Altena, W. F. 1999, AJ,
 117, 1792 (DGvA)
\reference Freeman, K. C. 1993, in ASP Conf. Ser. 48, The Globular
Cluster-Galaxy Connection, ed. G. H. Smith \& J. P. Brodie
(San Francisco: ASP), 608
\reference Gates, Gyuk, \& Turner, E. 1995, ApJ, 449, L123
\reference Gilmore, G., Wyse, R. F. G., \& Norris, J. E. 2002,
ApJ, 574, L39 (GWN)
\reference Gnedin, O. Y., Zhao, H.-S., Pringle, J. E., Fall, S. M.,
Livio, M., \& Meylan, G. 2002, ApJ, 568, L23
\reference Gould, A., Flynn, C., \& Bahcall, J. N. 1998, ApJ, 503, 798
\reference Ibata, R. A., Wyse, R. F. G., Gilmore, G., Irwin, M. J.,
\& Suntzeff, N. B. 1997, AJ, 113, 634
\reference Kinman, T. D., Cacciari, C., Bragaglia, A., Buzzoni, A.,
\& Spagna, A. 2003, in Galactic Dynamics Workshop, in press
(astro-ph/0211243) (K03)
\reference Majewski, S. R. 1992, ApJs, 78, 87
\reference Mizutani, A., Chiba, M., \& Sakamoto, T. 2003, ApJ, 589, L89
\reference Norris, J. E., \& Da Costa, G. S. 1995, ApJ, 447, 680
\reference Norris, J. E., Freeman, K. C., \& Mighell, K. L.1996,
ApJ, 462, 241
\reference Shetrone, M. D., C\^{o}t\'{e}, P., \& Sargent, W. L. W. 2001,
ApJ, 548, 592
\reference Tsuchiya, T., Dinescu, D. I., and Korchagin, V. I. 2003, ApJ, 589, L29
\reference White, S. D. M. 1983, ApJ, 274, 53
\reference Yamagata, T., \& Yoshii, Y. 1992, AJ, 103, 117
\reference Zaritsky, D., \& White, S. D. M. 1988, MNRAS, 235, 289

\end{document}